\def\selectedoptions{final}

\ifx\empty\selectedoptions
  \def\selectedoptions{final}
\fi

\documentclass[
   \selectedoptions
  ]
  {aipproc}

\def\selectedlayoutstyle {8x11double}
\layoutstyle\selectedlayoutstyle

\SetInternalRegister\hbadness{8000}

\newcommand\doingARLO[2][]{%
  \ifx\mmref\undefined #1\else #2\fi
}

\begin{document}

\def\apj{{\it Ap.\ J.,}}
\def\apjl{{\it Ap.\ J. Letters,}}
\def\aa{{\it Astron.\ Astrophys.,}}
\def\etal{{\it et al.}}
\def\annrev{{\it Ann.\ Rev.\ Astron.\ Ap.}}
\def\aplet{{\it Ap.\ Letters}}
\def\aj{{\it Astron.\ J.}}
\def\apj{Ap J}
\def\apjl{{\it Ap.J. Letters}}
\def\apjlet{{\it Ap.\ J.\ (Lett.)}}
\def\apjs{{\it Ap.\ J.\ Suppl.}}
\def\apjsup{{\it Ap.\ J.\ Suppl.}}
\def\aasup{{\it Astron.\ Astrophys.\ Suppl.}}
\def\aap{{\it Astron.\ Astrophys.\ Suppl.}}
\def\astro{{\it Astron.\ Astrophys.}}
\def\aa{{\it Astron.\ Astrophys.}}
\def\mnras{{\it M.\ N.\ R.\ A.\ S.}}
\def\iaucirc{{\it IAU Circular No .}}
\def\nature{{\it Nature}}
\def\nat{{\it Nature}}
\def\pasa{{\it Proc.\ Astr.\ Soc.\ Aust.}}
\def\pasp{{\it P.\ A.\ S.\ P.}}
\def\pasj{{\it PASJ}}
\def\pre{{\it Preprint}}
\def\qjras{{\it Quart.\ J.\ R.\ A.\ S.}}
\def\rppp{{\it Rep.\ ProAg.\ Phys.}}
\def\sovlet{{\it Sov. Astron. Lett.}}
\def\adspr{{\it Adv. Space. Res.}}
\def\expas{{\it Experimental Astron.}}
\def\ssr{{\it Space Sci. Rev.}}
\def\inpress{in press.}
\def\souspresse{sous presse.}
\def\inprep{in preparation.}
\def\enprep{en pr\'eparation.}
\def\submit{submitted.}
\def\soumis{soumis.}
\def\kte{kT$_{\rm e}$}
\def\ktbb{kT$_{\rm BB}$}
\def\rbb{R$_{\rm BB}$}
\def\ergs{ergs s$^{-1}$}
\def\erg{erg}
\def\ergscm{ergs s$^{-1}$ cm$^{-2}$}
\def\mdot{$\dot{\rm M}$}
\def\rin{R$_{\rm in}$}
\def\ledd{L$_{\rm Edd}$}
\def\comptt{{\sc Comptt}}
\def\msol{M$_\dot$}
\def\nh{N$_{\rm H}$}

\title 
      [FREGATE observation of a strong burst from SGR1900+14]
      {FREGATE observation of a strong burst from SGR1900+14}


\author{J-F. Olive}{
  address={Centre d'Etude Spatiale des Rayonnements, CNRS/UPS, 31028 Toulouse Cedex 04, France},
  email={olive@cesr.fr},
}

\author{K. Hurley}
{address={UC Berkeley Space Science Laboratory, Berkeley, CA 94720-7450}}

\author{J-P. Dezalay}{
  address={Centre d'Etude Spatiale des Rayonnements, CNRS/UPS, 31028 Toulouse Cedex 04, France},
}
\author{J-L. Atteia}{
  address={Centre d'Etude Spatiale des Rayonnements, CNRS/UPS, 31028 Toulouse Cedex 04, France},
}
\author{C. Barraud}{
  address={Centre d'Etude Spatiale des Rayonnements, CNRS/UPS, 31028 Toulouse Cedex 04, France},
}

\author{N. Butler}{
  address={Massachusetts Institute of Technology, Center for Space Research, Cambridge, MA, US},
}
\author{G. B. Crew}{
  address={Massachusetts Institute of Technology, Center for Space Research, Cambridge, MA, US},
}
\author{J. Doty}{
  address={Massachusetts Institute of Technology, Center for Space Research, Cambridge, MA, US},
}
\author{G. Ricker}{
  address={Massachusetts Institute of Technology, Center for Space Research, Cambridge, MA, US},
}
\author{R. Vanderspek}{
  address={Massachusetts Institute of Technology, Center for Space Research, Cambridge, MA, US},
}

% \copyrightholder{Acoustical Scociety of America}
\copyrightyear  {2001}

\begin{abstract}

After a long period of quiescence, the soft gamma repeater SGR1900+14
was suddenly reactivated on April 2001. On July 2$^{\rm nd}$ 2001, a
bright flare emitted by this source triggerred the WXM and FREGATE
instruments onboard the HETE-2 satellite. Unlike typical short
($\sim 0.1$ s) and spiky SGRs recurrent bursts, this event features a
4.1 s long main peak, with a sharp rise ($\sim$ 50 ms) and a slower
cutoff ($\sim$ 250 ms). This main peak is followed by a $\sim$ 2 sec
decreasing tail. We found no evidence of any precursor or any extended
`afterglow' tail to this burst. We present the preliminary spectral
fits of the total emission of this flare as observed by the FREGATE
instrument between 7 and 150 keV. The best fit is obtained with a
model consisting of two blackbody components of temperatures 4.15 keV
and 10.4 keV. A thermal bremsstrahlung can not be fitted to this
spectrum. We compare these features and the burst energetics with the
other strong or giant flares from SGR1900+14.

\end{abstract}

\date{\today}

\maketitle

\section{Introduction}

\ifthenelse{\equal\selectedlayoutstyle{6x9}}{\par\bfseries 
  Note: The entire paper will be reduced 15\% in the printing
  process. Please make sure all figures as well as the text within the
  figures are large enough in the manuscript to be readable in the
  finished book.\par\bfseries 
  Note: The entire paper will be reduced 15\% in the printing
  process. Please make sure all figures as well as the text within the
  figures are large enough in the manuscript to be readable in the
  finished book.\par\bfseries 
  Note: The entire paper will be reduced 15\% in the printing
  process. Please make sure all figures as well as the text within the
  figures are large enough in the manuscript to be readable in the
  finished book.\normalfont}{}

With only four (maybe five) objects known, the Soft Gamma Repeaters
(SGRs, see \cite{hur2000} for a review) are rare sources. They undergo
repeated, unpredictable periods of intense activity during which they
emit few hundreds of brief ($\sim 0.1$ s) and intense
($\sim10^{3}-10^{4}\ L_{Edd}$) bursts of soft $\gamma$-rays. Besides
these {\bf classic} bursts, very rarely, SGRs emit {\bf giant}
flares\footnote{So far, only two such giant flares have been detected
: the famous events of March 5 1979 from SGR0526-66 and August 27 1998
from SGR1900+14.}, which are extremely energetic (typically
$\sim10^{44}$ \erg) and much longer events, lasting for several
minutes. The SGR active phases usually last only from a few weeks to a
few months. They are separated by long periods (from years to decades)
of quiescence during which the SGRs are detected as persistent soft
X-ray sources associated with supernova remnants.

\begin{figure} [t!]
  \resizebox{1.7\columnwidth}{!}  {\includegraphics{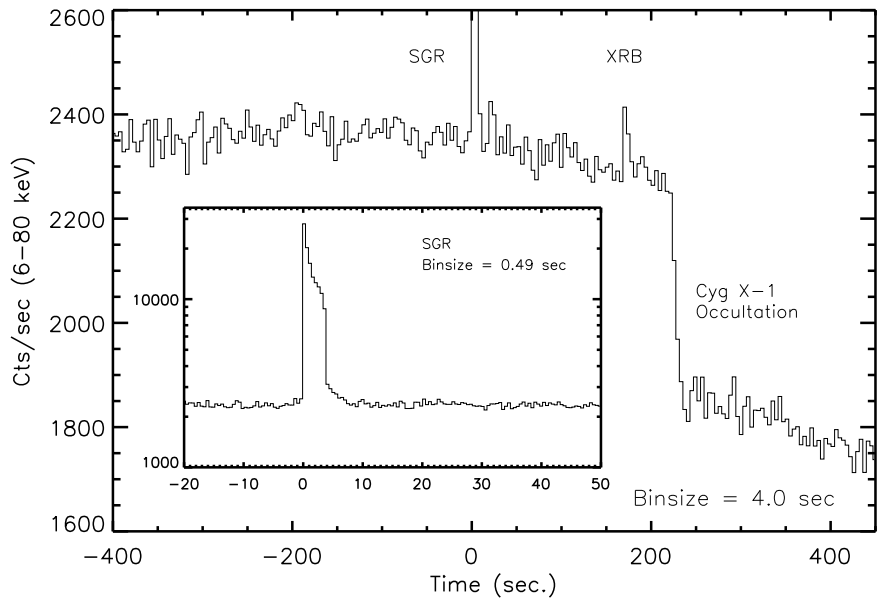}}
  \caption{The FREGATE 6--80 keV time history centered on the trigger
  time. The peak is cut off on the linear plot.} 
  \label{lc-th}
\end{figure}

Among the few Soft Gamma Repeaters, SGR1900+14 is a kind of
prototype. This source was first detected in 1979 (\cite{maz79}) when
it burst 3 times in 2 days. Its activity resumed in 1992
(\cite{kouv93}) and afterwards in May 1998 (\cite{kouv98},
\cite{hur99a}) and April 2001 (\cite{1041}, \cite{1043},
\cite{1045}). During the quiescence state, ASCA and RXTE observations
revealed a low luminosity ($\sim 3 \times 10^{34}$ \ergs) soft X-ray
source with a periodicity of 5.16~s and a high value of period
derivative\footnote{A similar spin period and rapid spindown is also
measured for SGR 1806-20 (\cite{kouv98b})} ($\sim 6 \times 10^{-11} \,
\rm s \, s^{-1}$, \cite{hur99b}, \cite{kouv99}, \cite{woo99}). 
SGR1900+14 lies just outside G42.8+0.6, a $10^{4}$-year-old
galactic supernova remnant. It is plausible that the SGR is a young
neutron star, born in this supernova. If the spindown of the neutron
star is due to magnetic braking, assuming purely dipole radiation, the
inferred magnetic field is found to be B $\rm \sim 8 \times 10^{14}
G$. All these indices support the hypothesis formulated in 1992 that
SGR are strongly magnetized young neutron stars (i.e. {\it magnetars}
\cite{DT92}). In a magnetar, the energy of the bursts are drawn from the
magnetic field energy which dominates all other sources of energy
including the neutron star rotation (\cite{TD95}).

The 2001 reactivation period of SGR1900+14 started with a burst
detected by {\it Ulysses} on April 17 (\cite{1045}) and a strong
$\sim$ 40 s flare detected by {\it BeppoSax} the day after
(\cite{1041}, \cite{1043}). In June-July 2001, SGR1900+14 was very
active and entered the antisolar field-of-view of the HETE
instruments. Many short {\bf classic} bursts were detected at that
time and are reported elsewhere. At 03:34:06.53 UTC on 2 July, the
HETE-2 FREGATE and WXM instruments detected and localized a strong
burst lasting 4.1 s from SGR1900+14 (\cite{1078}). In this paper, we
report on the preliminary timing and spectral analysis of this strong
burst as observed with FREGATE. The WXM data on this burst are also
presented in these proceedings (\cite{jap})

\begin{figure} [t!]
  \resizebox{1.1\columnwidth}{!}  {\includegraphics{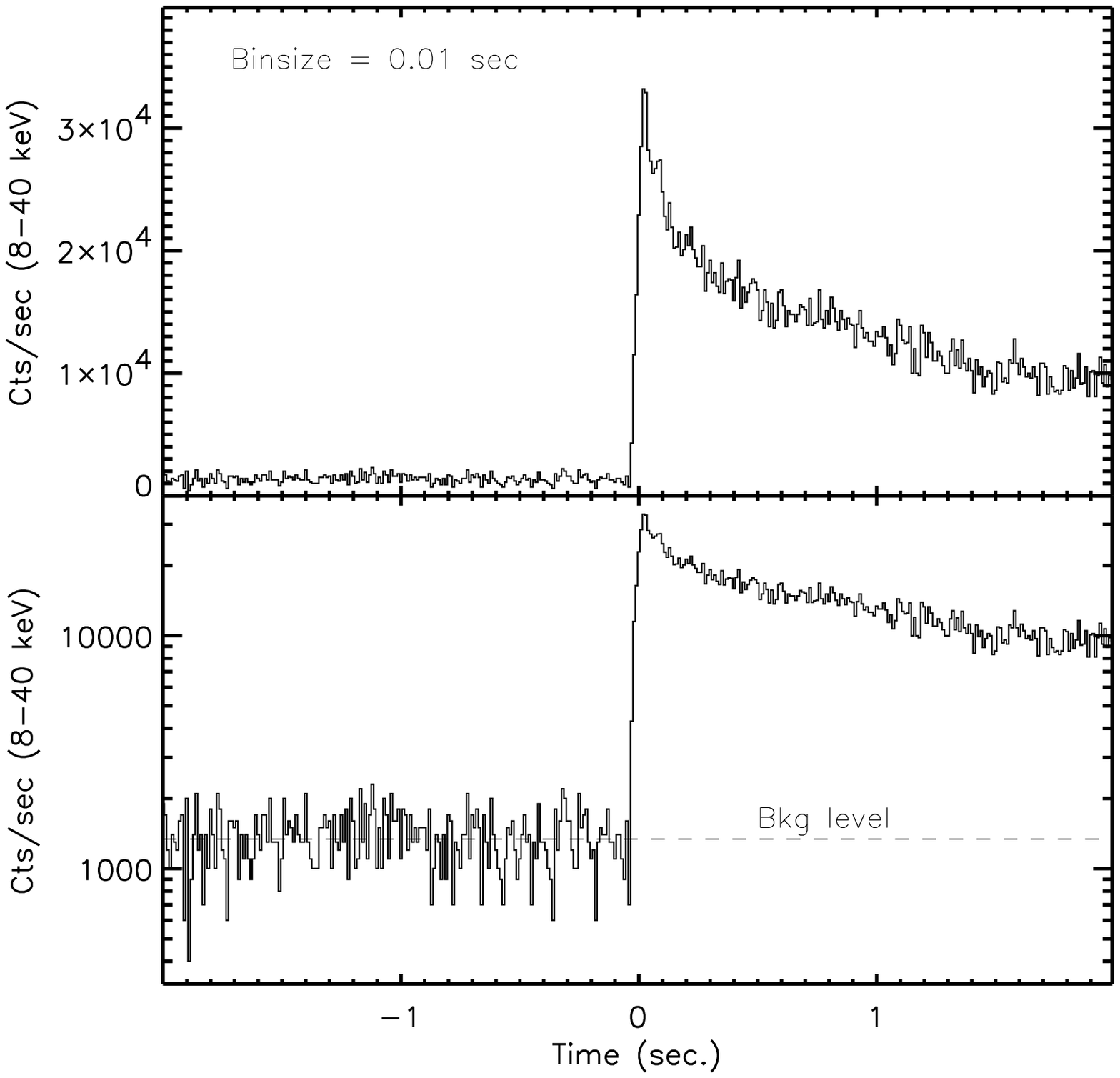}}
  \resizebox{1.1\columnwidth}{!}  {\includegraphics{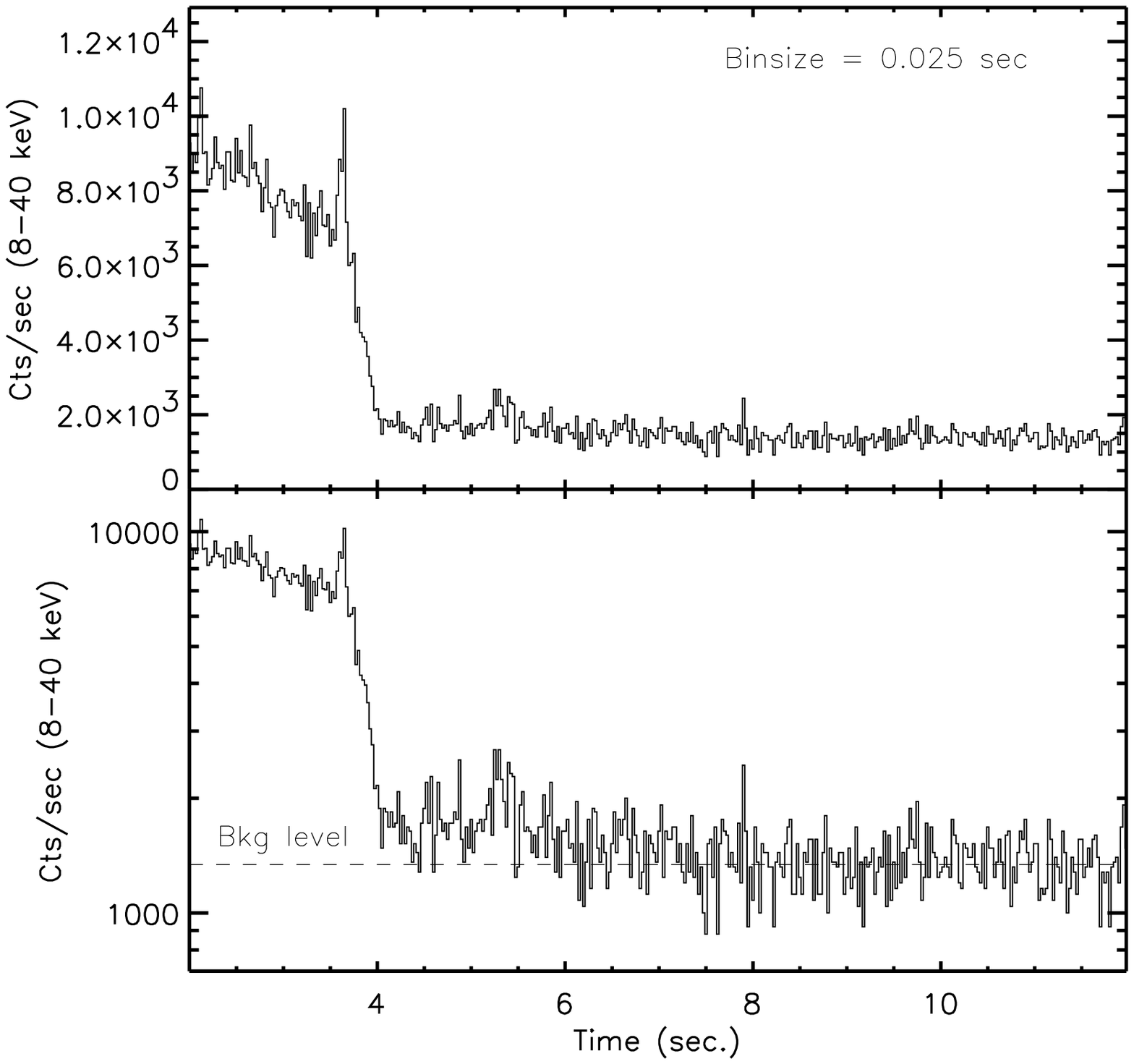}}
  \caption{The SGR flare light curves in 8-40 keV energy in a linear scale
  (top) and logarithmical scale (bottom)} \label{lc-sgr}
\end{figure}

\subsection{The FREGATE observations}

\subsubsection{Large scale timing analysis}

The four identical units of X/$\gamma$--ray detectors onboard HETE-2,
named FREGATE, are sensitive to photons between 6 and 400 keV (see
\cite{atteia-wh} for a full description of the experiment and
operating modes). Using the continuous time history data (resolution
0.16 s in 4 energy bands) we have searched indices for the SGR
activity before (i.e. precursors) and after (i.e. extended tail or
`afterglows') the main peak for which the instruments triggered. For
that purpose, we have built light curves in various energy bands and
with various time resolutions. We didn't find any such features.  As
an exemple, Figure \ref{lc-th} shows the FREGATE 6--80 keV time
history centered on the trigger time. On the large scale plot (more
than 10 minutes with 4 s resolution), we see that the count rate goes
back to the background level immediatly after the 4.1 s flare
(labelled `SGR' on the plot). Three other features to note: 170 s
after the trigger the very soft (i.e. no photons above 30 keV) and
short ($\sim$ 5 s) excess labelled `XRB' on the plot is probably due
to a galactic X-ray burster. Unfortunatly this burst was outside the
WXM field-of-view and was not localized. About four minutes after the
trigger, the large step visible in this light curve is due to an Earth
occultation of Cyg X-1. Finally, about 20 s after the SGR peak, the
slight increase in the count rate (marginally significant) is possibly
due to the variability of Cyg X-1 or to a background fluctuation.

\subsubsection{The SGR flare}

The inset in Figure \ref{lc-th} shows a zoom with a logarithmic scale
on the peak of the burst, with 0.49 s time resolution.  With the
exception of the main peak, the only noticeable feature is a short
$\sim 2-3$ s tail rapidly decreasing to the background level
after the peak.

A zoom on the first part of the burst is shown in Figure~\ref{lc-sgr}
(left). This light curve has been constructed with the `Burst Data' of
FREGATE (256k photons tagged in time with a resolution of 6.4 $\mu$s
and in 256 energy channels for each detector). The dashed line
represents the linear interpolation of the background level taken for
times less than 2 s before and greater than 10 s after the trigger
time. Several important features can be seen in this plot. First, no
short precursors are detected during the few seconds before the main
peak. The rise of the burst is quasi-linear in time and is fully
resolved ($\sim$ 50 ms). This sharp rise is followed by a short
($\sim$ 20 ms) spike occuring at the peak of the flare. Then the
intensity starts to decrease in a complex way. A zoom on the last part
of the burst is shown in Figure \ref{lc-sgr} (right). The main peak
terminates with a second short spike (lasting $\sim$ 100 ms)
immediatly followed by a rapid decrease of the intensity lasting for
$\sim 0.35 $ s. After the main peak, a short spiky tail lasting for
$\sim$ 2 s is visible.

\subsection{The spectrum and energetics of the flare}

As a first spectral analysis, we have built the spectrum of the whole
main peak (4.05 s duration) and a background spectrum for 10 s of data
before the burst using the `Burst Data' of FREGATE. The deconvolution
matrices and the energy-to-channel relations for that date have been
computed using the method validated with the Crab observations of
FREGATE (\cite{olive2002}). A 2 \% systematic error has been added to
the statistical errors to account for the calibration uncertainties
affecting this high-level spectrum. We have tried with XSPEC to fit
the 7-150 keV FREGATE spectrum with several simple spectral models
widely used for such bursts.

\begin{figure} [t!]
  \resizebox{1.95\columnwidth}{!}  
{\includegraphics{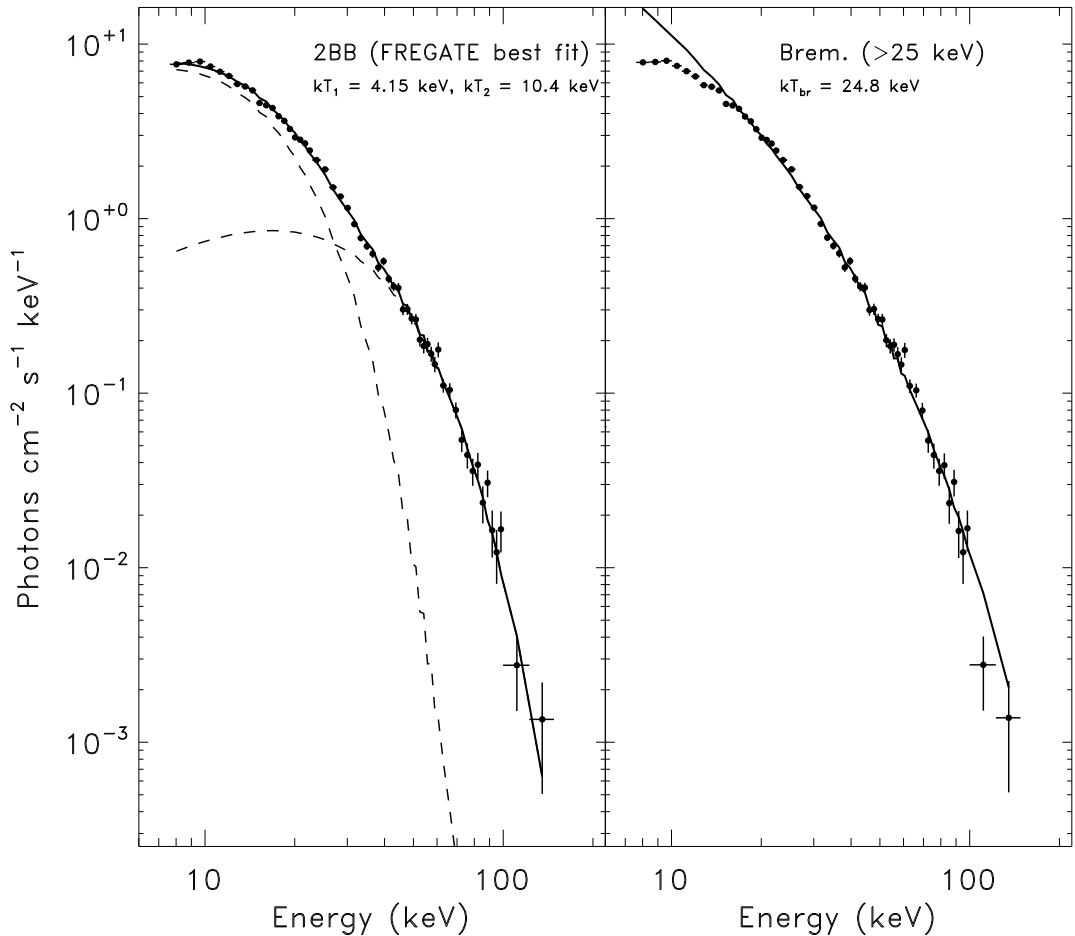}} 

 \caption{The unfolded total emission spectrum of the July 2 2001
 burst from SGR1900+14 seen by FREGATE. Right : a thermal
 bremsstrahlung above 25 keV can roughly reproduce the data but
 totally fails below $\sim$ 15 keV. Left : using the best fit
 composite model with 2 blackbody in the extended range 7--150 keV}

\label{ufs}
\end{figure}

During these trials, the spectrum readily appeared to feature two main
characteristics : (I) it is strongly curved below 20 keV, (II) it
extends up to 150 keV. None of the single component models (thermal
bremsstrahlung OTTB, powerlaw PL, blackbody BB, broken powerlaw,
powerlaw with an exponential cutoff, etc...) provides an acceptable
fit over the whole FREGATE range. As an example, we show in Figure
\ref{ufs} (right) the unfolded spectrum derived with an OTTB model for
energies above 25 keV. If the fit is marginally acceptable ($kT_{\rm
br}$ = 24.8 keV, $\chi^{2}$ = 1.49 for 47 dof), the model totally
fails to reproduce the low energy part of the spectrum below 15 keV
and the situation is even worse with the other simple models.

Among the composite models that we have tried (OTTB+BB, BB+PL, etc..)
only the sum of two blackbody components (2BB) produces a good fit
($\chi^{2}$ = 1.127 for 66 dof) over the full energy range (see Figure
\ref{ufs}, left). The temperatures we derived are kT${_1}$ = 4.15
$\pm$ 0.1 keV and kT${_2}$ = 10.4 $\pm$ 0.4 keV.

The flux of the burst was computed by extrapolating and integrating
the fitted 2BB model in the energy band $>$ 25 keV.  We have computed
the luminosity and the total energy for both spectral components and
the total emission assuming an isotropic emission at a source distance
of 10 kpc. All these results are reported in Table
\ref{tableparamtotal} and compared to those of 3 major bursts from the
source in the same energy range.

\section{Discussion and conclusions}

On August 27, 1998 {\it Konus-Wind}, {\it Ulysses}, and {\it BeppoSAX}
detected a {\bf giant} flare from SGR1900+14 (\cite{hur99a},
\cite{fer99}, \cite{maz99}). Its luminosity of $1.41 \times 10^{41}$
\ergs (above 25 keV) and exceptional duration ($\sim$ 6 min)
made this event the most intense burst ever detected at Earth,
thousands of times more energetic than the other bursts of this
source. The pulsations at 5.16 s were clearly detected during the
burst.

Only two days later (August 29), a strong bright burst was detected
simultaneously by BATSE and RXTE (named, after \cite{ibr01}, the {\bf
unusual} burst). This event exhibits a 3.5 s burst peak preceded by
complex ($\sim$ 1 s) precursor and followed by a long ($\sim$ 10$^{3}$
s) tail modulated at the 5.16 s pulsation period. Its luminosity of
$6.43 \times 10^{40}$ \ergs~is a factor of only 2 less than the giant
August 27 burst. Nevertheless, due to its shorter duration, the total
energy carried by this burst is much less by a factor $\sim$ 200.

\begin{table} [t!]
\begin{tabular}{rrccc}
\hline
    \tablehead{1}{r}{b}{Burst}
  & \tablehead{1}{c}{b}{Component}
  & \tablehead{1}{c}{b}{Flux \\ $10^{-6}$\ergscm}
  & \tablehead{1}{c}{b}{Luminosity \\ $10^{40}$\ergs}
  & \tablehead{1}{c}{b}{E$_{\rm tot}$ \\ $10^{40}$\erg} \\
\hline
Jul. 2$^{\rm nd}$ 2001 flare  &  4.15 keV BB~                      & 0.30 & 0.36 & 1.45   \\
                              &  10.4 keV BB~                      & 1.10 & 1.33 & 5.40   \\
                              &                                    &      &      &        \\
                              &   4.05 s total burst               & 1.40 & 1.69 & 6.85   \\
                              &                                    &      &      &        \\
\hline
 August 27, 1998 (giant)         &   $\sim$ 370 s burst               &      & 14.1 & 5200   \\
 August 29, 1998 (unusual)       &   $\sim$ 3.5 s burst               &      & 6.43 & 22.5   \\
 April 18, 2001 (intermediate)  &   $\sim$ 40 s burst                &      & 7.8  & 313   \\
\hline
\end{tabular}
\label{tableparamtotal}

 \caption{Summary for the energetics of the SGR1900+14 July 2$^{\rm
 nd}$ flare using the best spectral model (a sum of two blackbodies)
 in the range $> 25$ keV. For the calculations of the luminosity and
 total energy E$_{\rm tot}$ we assume a source distance of 10
 kpc. These parameters are compared to those of 3 major bursts from
 the source in the same energy range}

\end{table}

On April 18, 2001 {\it BeppoSAX} was triggered by an intense X-ray
burst from SGR1900+14 (\cite{1041}, \cite{1043}). The event, also
detected by {\it Ulysses}, lasted $\sim$ 40 s and was modulated with
the 5.16 s period. Unfortunatly, no spectral data are available for
this burst. Nevertheless, assuming an optically thin thermal
bremsstrahlung spectral model with $kT\sim$30 keV (\cite{iauc7611})
the inferred $25-100$ keV fluence is $\sim 6.5 \times 10^{-6}$
\ergscm~($\sim 7.8 \times 10^{40}$ \ergs~ at 10 kpc) that is
similar to the previous burst. Nevertheless, considering its
energetics and unusual duration, this event has been qualified as an
{\bf intermediate} burst.

The morphology of the July 2$^{\rm nd}$, 2001 flare as observed with
FREGATE resembles the August 29, 1998 {\bf unusual} flare. Its
duration is similar and the total energy carried by the burst above 25
keV is a factor of only 3--4 less. It is not surprising that we did
not detect any precursor or pulsed afterglow at the level reported
with {\it RXTE} (\cite{ibr01}). If we scale down these features by a
factor of $4 \times S_{\rm HETE}/S_{\rm RXTE} \sim 160$, they are
undetectable against the somewhat large FREGATE background due to its
extended field-of-view.

The main difference between the two bursts resides in the energy
spectrum of the main peak. The {\bf unusual} burst was so bright that
the {\it RXTE}-PCA detectors was saturated during the majority of the
peak.  Nevertheless, the burst rise and burst falloff spectra could be
fitted with a classical OTTB model\footnote{Alternatively a PL+BB
model also gives a good fit with kT $\sim 2.4-2.5$ keV and $\gamma
\sim 1.2-1.6$.} with temperatures of 17.2 $\pm$ 2 keV and 15.4 $\pm$
2.5 keV respectively. The BATSE spectrum of this burst could also be
fitted above 25 keV by an OTTB model with kT = $20.6 \pm 0.3$
keV. Both analyses combined suggest a nonvarying spectrum for the {\it
unsusual} burst (\cite{ibr01}).  The spectrum presented here can also
be fitted by an OTTB model {\bf above 25 keV} but it is much more
curved than this model at lower energies. There is no doubt that such
a feature would have been detected with {\it RXTE} if present for the
unusual burst. Tentatively we have tried to fit our spectrum with a
composite model consisting of two blackbodies with temperatures
kT${_1}$ = 4.15 $\pm$ 0.1 keV and kT${_2}$ = 10.4 $\pm$ 0.4 keV. The
equivalent radii for an isotropic emission are 25 and 3.5 km
respectively (at 10 kpc). These values are suggestive of emission
regions close to the stellar surface but our modelling is probably too
crude to draw any definite conclusion. Many effects, such as the
anisotropy of the heat flow through an ultramagnetized neutron star
envelope, the reprocessing by a light element atmosphere and the
general relativity correction can modify a thermal spectrum near a
magnetar surface leading to different values of temperatures and radii
(\cite{perna}).

A detailled discussion of the FREGATE observations of this burst in
the framework of the magnetar model is in preparation and will be
published in the near future (\cite{olive_futur}).

\doingARLO[\bibliographystyle{aipproc}]
          {\ifthenelse{\equal{\AIPcitestyleselect}{num}}
             {\bibliographystyle{arlonum}}
             {\bibliographystyle{arlobib}}
          }

%\bibliography{sample}

\end{document}